\documentclass[epsfig]{article}
\usepackage{amssymb}

\usepackage{graphicx}
\usepackage{amsmath}

\input epsf.sty
\newtheorem{theorem}{Theorem}
\newtheorem{acknowledgement}[theorem]{Acknowledgement}

\input{tcilatex}
\begin{document}

\title{\rightline{\mbox{\normalsize
{GNPHE/0503}}} \textbf{On Inflation Potentials in Randall-Sundrum} \textbf{\
Braneworld Model}}
\date{}
\author{M.Bennai$^{1,3}$\thanks{%
m.bennai@univh2m.ac.ma}, H.Chakir$^{2,3}$ and Z.Sakhi$^{2,3}$ \\
$^{\mathit{1}}$L.P.M.C, {\small Groupe de Physique des Hautes Energies et de
Cosmologie. Facult\'{e} des Sciences Ben M'sik, B.P. 7955, }\\
{\small \textit{Universit\'{e} Hassan II-Mohammedia,}} {\small Casablanca,
Maroc. }\\
$^{\mathit{2}}${\small \textit{LISRI, Facult\'{e} des Sciences Ben M'Sik,
Universit\'{e} Hassan II-Mohammedia, Casablanca,\ Maroc,} }\\
$^{3}${\small \ Groupement Nationale de Physique des Hautes Energies,  Point
focal, LabUFR-PHE, Rabat, Maroc.}}
\maketitle

\begin{abstract}
We study the inflationary dynamics of the universe in the Randall-Sundrum
typeII Braneworld model. We consider both an inverse-power law and
exponential potentials and apply the Slow-Roll approximation in high energy
limit to derive analytical expression of relevant inflationary quantities.
An upper bound for the coupling constant was also obtained and a numerical
value of \ perturbation spectrum is calculated in good agreement with
observation.


\end{abstract}

\newpage

\newpage

\section{Introduction}

\qquad Recently a great interest has been devoted to study cosmological
inflation \cite{inflation} in the framework of braneworld scenario; for a
review see \cite{braneworld}-\cite{RSII}. This issue has been motivated by
observations on accelerating universe \cite{accelerating} and dark energy 
\cite{darkenergy} as well as results on interpretations of these phenomena
in terms of scalar field dynamics \cite{scalar field,taychons}. In this
regards higher dimensional cosmological models have been built and new
solutions were obtained \cite{branesolut}. Of particular interest is the
Randall-Sundrum (RS) braneworld picture based on considering a type IIB
D3-brane embedded in AdS$_{5}$ \cite{RS}, where the conventional $4d$
gravity is recovered in low energy limit despite that fifth extra dimension
is non compact. As shown in \cite{RSII}, the RS braneworld model is
described by a $4d$ effective gravity induced on the world volume of the
D3-brane embedded in $5d$ Einstein gravity. The $5d$ Planck scale $M_{5}$ is
assumed considerably smaller than the corresponding $4d$ effective Planck
scale, $M_{4}=1.2\times 10^{19}GeV$ opening an issue for new perspectives in
cosmology at low energies. Since RS development, diverses inspired RS brane
inflationary cosmological models have been constructed \cite{Maartens,exp}
and new insights have been gained. Like in standard inflation, most of these
cosmological models rest on a single scalar field rolling in a given
inflaton field potential $V\left( \phi \right) $.

In this paper, we consider inflation dynamics in braneworld cosmology, first
with an inverse-power low potential $V\left( \phi \right) =\mathrm{\mu }%
^{\alpha +4}/\phi ^{\alpha }$ and then with an exponential one; namely $%
V(\phi )=V_{0}e^{-\beta \phi }$; the $\mathrm{\mu },\alpha $ and $\beta $
moduli will be computed later on. These two types of potential are
interesting in cosmology since they are used to modeling tachyonic and
quintessential matter \cite{taychons,liddle}. Using slow roll approximation,
we compute, in the high energy limit, the inflaton time evolution $\phi
=\phi \left( t\right) $ and determine the corresponding scale factor $%
a=a\left( t\right) $ and other cosmological quantities. We also derive an
upper bound for the coupling constant and give a numerical value of
perturbation spectrum, which is in good agreement with observation.

The presentation of this work is as follows: In section 2, we present the
basic equations of braneworld inflation assuming the Randall-Sundrum's
second model, and recall some known results, especially for chaotic
inflation. In section 3, we present our results on scalar field dynamics in
braneworld inflation, for both inverse power law and exponential potentials.
In this context, we derive the time evolution of\ the scalar field and scale
factor. Various spectral quantities are also calculated and an upper value
of coupling constant is obtained. Our results are compared to those obtained
in standard inflationary cosmology and are shown to be in good agreement
with the observable quantities \cite{observ}. We end this paper by a
conclusion.

\section{Inflation in Randall-Sundrum \textbf{braneworld} model}

\subsection{\protect\bigskip Modified Einstein-Friedmann equations}

We start this section by recalling briefly some fundamentals on
Randall-Sundrum type II braneworld model\cite{RSII}. We first give the
braneworld cosmological Einstein-Friedmann equations; then we discuss
resulting modifications of Friedmann equations, slow roll parameters and
perturbation spectrum.

Starting from $5d$ Einstein equation with cosmological constant $\Lambda $,
and by supposing that matter fields is confined on D3-brane, Shiromizu 
\textit{et al}.\cite{Shiromizu} have shown that 4d Einstein equation induced
on the brane reads as 
\begin{equation}
G_{\mu \nu }=-\Lambda _{4}g_{\mu \nu }+\left( {\frac{8\pi }{M_{4}^{2}}}%
\right) T_{\mu \nu }+\left( {\frac{8\pi }{M_{5}^{3}}}\right) ^{2}\pi _{\mu
\nu }-E_{\mu \nu }.  \label{1}
\end{equation}
In this relation, $T_{\mu \nu }$ is the energy-momentum tensor of matter on
the brane, $\pi _{\mu \nu }$ is a tensor quadratic in $T_{\mu \nu }$ and $%
E_{\mu \nu }$ is a projection of the five-dimensional Weyl tensor describing
the effect of bulk graviton degrees of freedom on brane dynamics. The
effective cosmological constant $\Lambda _{4}$ on the brane is determined by
the five-dimensional bulk cosmological constant $\Lambda $ and the 3-brane
tension $\lambda $ as shown below, 
\begin{equation}
\Lambda _{4}={\frac{4\pi }{M_{5}^{3}}}\left( \Lambda +{\frac{4\pi }{%
3M_{5}^{3}}}\lambda ^{2}\right) .  \label{2}
\end{equation}
Recall also that $4d$ and $5d$ Planck scales $M_{4}$ and $M_{5}$\ are
related as, 
\begin{equation}
M_{4}=\sqrt{{\frac{3}{4\pi }}}\left( {\frac{M_{5}^{2}}{\sqrt{\lambda }}}%
\right) M_{5},  \label{3}
\end{equation}
with $\lambda $\ as before. In braneworld cosmological model where the
metric projected onto the brane is a spatially flat
Friedmann-Robertson-Walker model with scale factor $a(t)$, Friedmann
equation on the brane reads as~\cite{bdel}, 
\begin{equation}
H^{2}={\frac{\Lambda _{4}}{3}}+\left( {\frac{8\pi }{3M_{4}^{2}}}\right) \rho
+\left( {\frac{4\pi }{3M_{5}^{3}}}\right) ^{2}\rho ^{2}+{\frac{\mathcal{E}}{%
a^{4}}},  \label{4}
\end{equation}
where $\mathcal{E}$ is an integration constant arising from $E_{\mu \nu }$
and thus transmitting bulk graviton influence onto the brane. This term
appears as a form of ``dark radiation''\ and may be fixed by observation 
\cite{bdel,dr}. However, during inflation this term is rapidly diluted, so
we will neglect it. We will also assume that, in the early universe, the
bulk cosmological constant is $\Lambda \sim -4\pi \lambda ^{2}/3M_{5}^{3}$
so that $\Lambda _{4}$ is negligible. With above assumptions, braneworld
Friedmann equation(\ref{4}) reduces to, 
\begin{equation}
H^{2}={\frac{8\pi }{3M_{4}^{2}}}\rho \left[ 1+{\frac{\rho }{2\lambda }}%
\right] .  \label{5}
\end{equation}
Note that the crucial correction to standard inflation is given by the
density quadratic term $\rho ^{2}$. Brane effect is then carried here by\
the deviation factor ${\frac{\rho }{2\lambda }}$ with respect to unity. This
deviation has the effect of modifying the dynamics of the universe for
densities $\rho \gtrsim \lambda $. Note also that in the limit $\lambda
\rightarrow \infty ,$ we recover standard four-dimensional general
relativistic results (neglecting $\mathcal{E}$). Note moreover that in
inflationary theory with inflaton potential $V(\phi )$, energy density $\rho
=\rho \left( \phi \right) $ and pressure $p=p\left( \phi \right) $ are given
by the following relations,

\begin{equation}
\rho =\frac{1}{2}\dot{\phi}^{2}+V(\phi );\qquad p=\frac{1}{2}\dot{\phi}%
^{2}-V(\phi ),  \label{6}
\end{equation}
where $\phi $ is the inflaton field and the dot stands for the derivative
with respect time $t$. The potential $V(\phi )$ is the initial vacum energy
responsible of inflation. Along with these equations, one also has the
second inflation Klein-Gordon equation governing the dynamic of the scalar
field $\phi $ 
\begin{equation}
\ddot{\phi}+3H\dot{\phi}+V^{\prime }=0.  \label{7}
\end{equation}
This is a second order evolution equation which follows from conservation
condition of energy-momentum tensor $T_{\mu \nu }$ on 3-brane dominated by a
scalar field $\phi $ with a self-interaction potential $V(\phi )$. To
calculate physical quantities like scale factor or perturbation spectrum,
one has to solve eqs(\ref{5},\ref{7}) for some specific potentials $V(\phi )$%
. To do so, the following approximations are needed.

\subsection{Slow-roll approximation and perturbation spectrum on brane}

Inflationary dynamics requires that inflaton field $\phi $ driving inflation
moves away from the false vacuum and slowly rolls down to the minimum of its
effective potential $V(\phi )$ \cite{LINDE}. In this scenario, the initial
value $\phi _{i}=\phi \left( t_{i}\right) $ of the inflaton field and the
Hubble parameter $H$ are supposed large and the scale factor $a\left(
t\right) $ of the universe growth rapidly. Using Friedman equation, the
inflation condition $\ddot{a}>0$ allows us to derive the following bound on
pressure, 
\begin{equation}
\ddot{a}>0\qquad \Rightarrow \qquad p<-{\frac{\lambda +2\rho }{3\left(
\lambda +\rho \right) }\rho }.  \label{8}
\end{equation}
In the limit $\rho /\lambda \rightarrow \infty $, this condition reduces to$%
\ p<-{\frac{2}{3}}\rho $; this is a more restrictive constraint relation
than the corresponding one in standard inflation relation which requires $p<-%
\frac{\rho }{3}$. Applying the slow roll approximation, $\dot{\phi}^{2}\ll V$
and $\ddot{\phi}\ll V^{\prime },$ to brane field equations (\ref{5},\ref{7}%
), we obtain: 
\begin{equation}
H^{2}\simeq {\frac{8\pi V}{3M_{4}^{2}}}\left( 1+{\frac{V}{2\lambda }}\right)
\,,\qquad \dot{\phi}\simeq -{\frac{V^{\prime }}{3H}}.  \label{9}
\end{equation}
The presence of the factor ($1+{\frac{V}{2\lambda })}$ carries the
brane-modification with respect to the standard slow-roll expression
recovered by taking the limit $\lambda \rightarrow \infty $. Note that slow
roll approximation puts a constraint on the slope and the curvature of the
potential; this is clearly\ seen on the field expressions of $\epsilon $ and 
$\eta $ parameters given by \cite{Maartens},\ 
\begin{eqnarray}
\epsilon &=&-\frac{\overset{\cdot }{H}}{H^{2}}\equiv {\frac{M_{4}^{2}}{4\pi }%
}\left( {\frac{V^{\prime }}{V}}\right) ^{2}\left[ \frac{\lambda (\lambda +V)%
}{(2\lambda +V)^{2}}\right] ,  \label{10} \\
\eta &=&\frac{V^{\prime \prime }}{3H^{2}}\equiv {\frac{M_{4}^{2}}{4\pi }}%
\left( {\frac{V^{\prime \prime }}{V}}\right) ^{2}\left[ \frac{\lambda }{%
2\lambda +V}\right] .  \label{11}
\end{eqnarray}
Slow-roll approximation takes place if these parameters are such that $%
\mathrm{max}\{\epsilon ,|\eta |\}\ll 1$ and inflationary phase ends when $%
\epsilon $ and $\left| \eta \right| $ are equal to one. The other important
quantity related to inflation is the number $N$ of e-folding, indicating the
growing of the size of universe. Using slow roll approximation, this $N$
number reads in present case as follows,

\begin{equation}
N\simeq -{\frac{8\pi }{M_{4}^{2}}}\int_{\phi _{\mathrm{i}}}^{\phi _{\mathrm{f%
}}}{\frac{V}{V^{\prime }}}\left( 1+{\frac{V}{2\lambda }}\right) d\phi .
\label{12}
\end{equation}
Before proceeding it is interesting to comment low and high energy limits of
these parameters. Note that at low energies where $V\ll \lambda $, the
slow-roll parameters take the standard form. At high energies $V\gg \lambda $%
, the extra contribution to the Hubble expansion dominates and the new
factors in square brackets of eqs(\ref{10}-\ref{11}) become of order $%
\lambda /V$. The number of e-folding in the limit $V\gg \lambda $, 
\begin{equation}
N\simeq -\frac{4\pi }{\lambda M_{4}^{2}}\int_{\phi _{i}}^{\phi _{f}}\frac{%
V^{2}}{V^{\prime }}d\phi ,  \label{13}
\end{equation}
where $\phi _{i}$ and $\phi _{f}$ stand for initial and final value of
inflaton.

\qquad To test inflation model, one must compute the spectrum of
perturbations produced by quantum fluctuations of fields around their
homogeneous background values. Using slow-roll equations and following \cite
{Maartens}, the scalar amplitude $A_{\QTR{sc}{s}}^{2}$ of density
perturbation, evaluated by neglecting back-reaction due to metric
fluctuation in fifth dimension ($E_{\mu \nu }=0)$, is given by 
\begin{equation}
A_{\QTR{sc}{s}}^{2}\simeq \left. \left( {\frac{512\pi }{75M_{4}^{6}}}\right) 
{\frac{V^{3}}{V^{\prime 2}}}\left[ {\frac{2\lambda +V}{2\lambda }}\right]
^{3}\right| _{k=aH}.  \label{14}
\end{equation}
Note that for a given positive potential, the $A_{\QTR{sc}{s}}^{2}$
amplitude is increased in comparison with the standard result. Note also
that in high energy limit this quantity behaves as, 
\begin{equation}
A_{S}^{2}\simeq \frac{64\pi }{75\lambda ^{3}M_{4}^{6}}\frac{V^{6}}{%
V^{^{\prime 2}}}.  \label{15}
\end{equation}
On the other hand, using eqs(\ref{10}-\ref{11}), one can compute the
perturbation scale-dependence described by the spectral index $n_{\QTR{sc}{s}%
}\equiv 1+d\left( \ln A_{\QTR{sc}{s}}^{2}\right) /d\left( \ln k\right) $. We
find, 
\begin{equation}
n_{\QTR{sc}{s}}-1\simeq 2\eta -6\epsilon \,,  \label{16}
\end{equation}
Note that at high energies $\lambda /V$ , the slow-roll parameters are both
suppressed; and the spectral index is driven towards the Harrison-Zel'dovich
spectrum, $n_{\QTR{sc}{s}}\rightarrow 1$ as $V/\lambda \rightarrow \infty $.

In what follows, we shall apply the braneworld formalism that we have
described above by singling out two specific kinds of inflaton potentials.
These are the inverse power law potential and the exponential one that have
gained revival interest in recent literature in connection with dark matter
and quintessence cosmology \cite{liddle} .

\section{Scalar Field dynamics in braneworld scenario}

\qquad To begin, recall that chaotic inflationary model, which was first
introduced by Linde \cite{LINDE}, has been reconsidered recently by several
authors in the context of braneworld scenario \cite{Maartens,Paul}. In
present work, we are interested by coupling constants for inverse power-law
potential $V\left( \phi \right) =\frac{\mathrm{\mu }^{\alpha +4}}{\phi
^{\alpha }}$ and exponential one $V\left( \phi \right) =V_{0}\exp (-\beta
\phi $).

\subsection{Inverse-Power Law potential}

In the braneworld cosmology high energy limit where $V\gg \lambda $, brane
effect becomes important and the Friedmann equations are simplified as,

\begin{equation}
H^{2}\simeq \,\left( {\frac{4\pi }{3M_{5}^{3}}}\right) V^{2},\qquad \dot{\phi%
}\simeq -{\frac{V^{\prime }}{V}}\left( {\frac{M_{5}^{3}}{4\pi }}\right) .
\label{20}
\end{equation}
and Slow-Roll parameters (\ref{10}-\ref{11}) reduces to:

\begin{equation}
\epsilon \equiv {\frac{M_{4}^{2}}{16\pi }}\left( {\frac{V^{\prime }}{V}}%
\right) ^{2}\left[ {\frac{4\lambda }{V}}\right] \,,\qquad \eta \equiv {\frac{%
M_{4}^{2}}{8\pi }}\left( {\frac{V^{\prime \prime }}{V}}\right) \left[ {\frac{%
2\lambda }{V}}\right] .  \label{21}
\end{equation}
Consider the inverse power law potentiel given by, 
\begin{equation}
V=\frac{\mathrm{\mu }^{\alpha +4}}{\phi ^{\alpha }},  \label{22}
\end{equation}
where $\mathrm{\mu }$ is the inflaton coupling constant and $\alpha $ some
critical exponent. This potential has been studied in various occasions; in
particular in connection with quintessence in brane inflation \cite{liddle}
and tachyonic inflation \cite{tachyon1}. One of the interesting results in
this matter, and which is due to Huey and Lidsey\cite{liddle}, is that
inflation is generated for the range $\alpha >2$. Combining results on
universe observation and Slow-Roll approximation, we want to show that is
possible to express\ the inverse power law potentiel coupling constant $%
\mathrm{\mu }$ in terms of $\alpha $\ and $M_{5}$. To that purpose, consider
the field expression of the Slow-Roll parameter $\epsilon $ namely, 
\begin{equation}
\epsilon \simeq \frac{3M_{5}^{6}}{16\pi ^{2}}\left( \frac{\alpha ^{2}}{%
\mathrm{\mu }^{\alpha +4}}\right) \frac{1}{\phi ^{2-\alpha }}.  \label{23}
\end{equation}
Then using the constraint relation $\epsilon \sim 1$ characterizing end of
inflation, one can invert above relation as follws, 
\begin{equation}
\phi _{end}^{2-\alpha }=\frac{3M_{5}^{6}\alpha ^{2}}{16\pi ^{2}\mathrm{\mu }%
^{\alpha +4}}.  \label{24}
\end{equation}
Similarly, we can compute the e-folding number by help of eq(\ref{13}). We
find, 
\begin{equation}
N\simeq \frac{16\pi ^{2}\mu ^{\alpha +4}}{3M_{5}^{6}}\frac{1}{\alpha
(2-\alpha )}\left[ \phi _{f}^{2-\alpha }-\phi _{i}^{2-\alpha }\right] ,
\label{25}
\end{equation}
where $\phi _{i}$ and $\phi _{f}$\ stand for initial and end inflaton field
values. From these equations (\ref{24}-\ref{25}), we can deduce the
expression of $\phi _{i}$ in term of the e-folding number $N$, 
\begin{equation}
\phi _{i}^{2-\alpha }=\frac{3M_{5}^{6}\alpha }{16\pi ^{2}\mathrm{\mu }%
^{\alpha +4}}\left[ \alpha -N\left( 2-\alpha \right) \right] .  \label{26}
\end{equation}
Now, using\ observation data giving $N_{cobe}\approx 55$\cite{N} and
identifying the initial and final inflaton field values of inflation
interval as $\phi _{i}=\phi _{cobe}$ and $\phi _{f}=\phi _{end}$, we get,

\begin{equation}
\phi _{cobe}^{2-\alpha }=\frac{3M_{5}^{6}\alpha }{16\pi ^{2}\mu ^{\alpha +4}}%
\left[ 56\alpha -110\right] .  \label{27}
\end{equation}
Putting back into eq(\ref{15}) after substituting equation (\ref{22}), we
deduce the following expression of scalar amplitude, 
\begin{equation}
A_{S}\simeq \frac{64\pi (\mathrm{\mu }^{\alpha +4})^{2}}{45M_{5}^{9}\alpha }%
\phi _{cobe}^{1-2\alpha }.  \label{28}
\end{equation}
Using the observed numerical value of $A_{S}$ from COBE namely $A_{S}\sim
2.10^{-5}$, we can invert above identity to fix the inflaton coupling
constant $\mathrm{\mu }$ in term of $5d$ Planck mass $M_{5}$ and inflaton
field exponent $\alpha $ as shown below, 
\begin{equation}
\mathrm{\mu }^{\alpha +4}=\left[ \frac{90.10^{-5}\alpha M_{5}^{9}}{64\pi }%
\right] ^{\frac{2-\alpha }{3}}\left[ \frac{8\pi ^{2}}{\alpha
M_{5}^{6}(84\alpha -165)}\right] ^{\frac{1-2\alpha }{3}}.  \label{29}
\end{equation}
Note that to get sufficient inflation $(N=55)$ one can obtain from
eqs(24,26) a constraint on the initial value of the field wich depends on $%
\alpha .$ For $\alpha =4$, for example, we get $\phi _{i}<162.2$ $M_{4}.$

Following the same method, it is not difficult to check, by help of eqs(\ref
{11},\ref{22}), that the spectral index $n_{S}$ reads as,

\begin{equation}
n_{S}-1\simeq -6\epsilon +2\eta =\frac{1-2\alpha }{28\alpha -55},  \label{30}
\end{equation}
or equivalently $n_{S}=\frac{26\alpha -55}{28\alpha -55}$ whose positivity
condition requires $\alpha >2+\frac{3}{26}$ in agreement with Huey and
Lidsey prediction \cite{liddle}. Fixing a value of the inflaton field
exponent $\alpha $, we can compute the inflaton field coupling constant $%
\mathrm{\mu }$ and the spectral index $n_{S}$. For $\alpha =3$ for instance,
we get $n_{S}=0.83$ and for large $\alpha $, we have $0.92$ in perfect
agreement with observation \cite{N} In what follows, we consider the case of
inflation exponential potential.

\subsection{Exponential inflation on brane}

Here we consider our second brane inflation model with an exponential
potential \cite{exponential}. This potential has been used to study
tachyonic inflation \cite{taychons} and quintessence \cite{q}. In the
present work, we give a quantitative study and compute physical quantities
relevant to observation; in particular the perturbation spectrum. To that
purpose consider the exponential potential, 
\begin{equation}
V(\phi )=V_{0}e^{-\mathrm{\beta }\phi }  \label{31}
\end{equation}
where $V_{0}$ is a constant\ and $\mathrm{\beta }$ is the coupling strength
of the scalar field. By integrating eqs(\ref{9}), one can derive the time
evolution of scalar field. We find $\phi (t)=C+\frac{M_{5}^{3}}{4\pi }%
\mathrm{\beta }t,$ where\ $C$ is an integration constant. To obtain the
scale factor $a(t)$, one have to integrate the Friedmann equation (\ref{20}%
); we get,

\begin{equation}
a(t)=a_{i}\exp \left( -\frac{V_{0}}{3}\frac{16\pi ^{2}}{\mathrm{\beta }%
^{2}M_{5}^{6}}\exp \left[ -\mathrm{\beta }(C+M_{5}^{3}\mathrm{\beta }t)%
\right] \right) .  \label{32}
\end{equation}
From the study of the inflection point of this relation, one may compute $%
t_{end},$ the end time of inflation; this is given by

\begin{equation}
t_{end}=\frac{4\pi }{\mathrm{\beta }^{2}M_{5}^{3}}\left[ -\mathrm{\beta }%
C-\ln \left( \frac{3M\mathrm{\beta }_{5}^{6}}{16\pi ^{2}V_{0}}\right) \right]
.
\end{equation}
Putting back into $\phi _{end}(t)=C+\frac{M_{5}^{3}}{4\pi }\mathrm{\beta }%
t_{end}$, we determine the value of the scalar field at the end of inflation,

\begin{equation}
\phi _{end}=\frac{1}{\mathrm{\beta }}\ln \left( \frac{16\pi ^{2}V_{0}}{3%
\mathrm{\beta }^{2}M_{5}^{6}}\right) .
\end{equation}
Using eqs(\ref{13}), we can compute the number $N$ of\ e-folding for the
exponential potential, 
\begin{equation}
N=-\frac{16\pi ^{2}V_{0}}{3\mathrm{\beta }^{2}M_{5}^{6}}\left( e^{-\mathrm{%
\beta }\phi _{f}}-e^{-\mathrm{\beta }\phi _{i}}\right) ,
\end{equation}
where $\phi _{i}$ is the value of the scalar field at the beginning of
inflation, 
\begin{equation}
\phi _{i}\simeq -\frac{1}{\mathrm{\beta }}\ln \left( \frac{21\mathrm{\beta }%
^{2}M_{5}^{6}}{2\pi ^{2}V_{0}}\right) .
\end{equation}

We assume now that the number of e foldings before the end of inflation at
which observable perturbations are generated corresponds to $N=55$\cite{N}
and setting $\phi _{f}=\phi _{end},$ $\phi _{i}=\phi _{cobe}$, and $%
A_{S}^{2}=2.10^{-5}$, we can give an estimation of the coupling constant $%
\mathrm{\beta }$ of the scalar field. Straightforward computation leads to, 
\begin{equation}
\mathrm{\beta }=1.07\times 10^{-2}M_{5}^{-1}.
\end{equation}
Since $M_{5}<M_{4},$ one can deduce an upper bound limit of the coupling
constant $\mathrm{\beta }$ as shown below, 
\begin{equation}
\mathrm{\beta }>0.877\times 10^{-21}
\end{equation}
This will give a constraint on the initial value of the scalar field and
should be compared with standard inflation result for same potential for
which $\mathrm{\beta }_{0}=708,9\times 10^{-2}M_{4}^{-1}$. Note that like
for chaotic inflation, we have here also a very weakly coupled scalar field
on the brane compared to that of standard inflation. Using high energy limit
and following same analysis as for inverse power law potential, we can
compute spectral index $n_{s}$ for the exponential case. We find,\ 
\begin{equation}
n_{_{S}}=0.92
\end{equation}
which is in good agreement with observation for wich 
\begin{equation}
0.8<n_{s}<1.05.
\end{equation}
As far as this result is concerned, it is interesting to note that a similar
value was also obtained in \cite{sen} for tachyonic inflation.

\section{Conclusion}

In this paper, we have studied aspects of inflationary dynamics in the
framework of braneworld cosmology. It has been shown that in this scenario,
the Friedmann equations governing the dynamics of scalar field in the
universe are modified by an extra term which depend on the ratio of the
density of matter $\rho $ and the brane tension $\lambda \QTR{sl}{.}$\textsl{%
\ }In Slow-Roll approximation, this ratio depend only on the potential V and
the tension $\lambda .$

In this short letter, we have studied brane effects on inflation for both
the inverse power low and exponential potentials. In this context, we have
calculated the scale factor for the potential(\ref{31}) in braneworld
formalism and shown that it has an exponential form, as it should be in
inflation theory. We have also calculated a numerical value of the
perturbation spectrum for potentials (\ref{22},\ref{31}) in good agreement
with observation and an upper bound for the coupling constant for
exponential potential was obtained. This work may be applied to a concret
physical problems such as tachyonic inflation and dark matter.

\begin{acknowledgement}
The authors thank Prof. E.H.Saidi for helpfull discussion
\end{acknowledgement}

\end{document}